**Estimated scale of Born rule violation in superconducting qubit measurement**

**Jonathan F. Schonfeld**

Center for Astrophysics | Harvard and Smithsonian

60 Garden St., Cambridge MA 02138 USA

jschonfeld@cfa.harvard.edu

ORCID ID# 0000-0002-8909-2401

**Abstract:** I estimate the size of expected Born rule violation for a two-level superconducting qubit with dispersive readout. The estimate is based on extrapolating from an earlier analysis of experimental data on cloud chamber detection. That analysis made no explicit use of quantum measurement axioms and found indication that the Born rule breaks down when it would otherwise naively predict extremely small measurement probabilities. In such cases, the Born rule significantly over-predicts the number of measurement events. The level of breakdown that I estimate here for the two-level qubit may be too small to have a meaningful impact on practical quantum computing.

**Keywords:** Quantum measurement problem; Born rule; cloud chamber; qubit

**1. Introduction**

The Born rule is a cornerstone of quantum computing because it prescribes the essential connection between wavefunctions as abstract entities and phenomena that people or machines actually observe and count [1]. It has the status of an axiom, but has always been under suspicion because the rule makes no explicit reference to how specific measurements or observations are actually designed. Over the last several years [2-6], I have studied the detailed microphysics of real quantum measurement systems and found evidence indicating that the Born rule, rather than fundamental or exact, is actually emergent and approximate. In reference [3], I attempted to characterize where the Born rule breaks down for charged particle detection in a cloud chamber, using opportunistic data publicly available on the Internet. In the present paper, I attempt to extend that result to a two-level superconducting qubit, which is much closer to the engineering reality of quantum computing. As we shall see, violation of the Born rule in this case may be too small for meaningful impact on conceivable real-world quantum computation.

The next section reviews the microscopic theory and phenomenology of the Born rule for nuclear decay in a cloud chamber. This will establish the basic notation and mathematical machinery for scaling to the qubit in Section 3. Section 4 contains concluding remarks.

**2. Cloud chamber theory and phenomenology**

A cloud chamber is a container of air supersaturated with a vapor, typically of alcohol [6]. Vapor molecules typically form clusters, either randomly due to thermal fluctuations, or nucleated by a passing charged particle when it ionizes an air or vapor molecule. When an atomic nucleus that undergoes s-wave alpha decay is inserted into a cloud chamber, one observes a single track of alpha-induced vapor droplets moving in a single direction away from the nucleus, even though the initial alpha wavefunction must be spherically symmetric. A *typical* track droplet is nucleated by the passing alpha particle ionizing an air molecule. The *first* droplet – which determines the specific track direction – is nucleated when there is an exceptional vapor cluster that forms randomly with extremely large ionization cross section. Indeed, the exceptional cross section is so large that essentially the entire initial spherically symmetric alpha wavefunction is captured into a collimated beam emanating from the exceptional cluster. The cross section is very large because the energy of polarization that ionization induces in the cluster nearly compensates for the binding energy of the ejected electron, which is a singular condition in inelastic Coulomb scattering. From this picture, one can derive a Born rule for the probability distribution of the locations at which tracks originate. There should be a limit to this Born rule because there should be a limit to how well induced polarization can compensate for electron binding energy. And, in turn, there should be a limit on such compensation because a vapor cluster is a stick figure of finitely many discrete molecules, rather than a droplet of a continuous, indefinitely adjustable medium (although the calculation below indicates that the minimum possible increment, while nonzero, is extremely small). In [3] I identified evidence suggesting Born rule breakdown in a cloud chamber. Breakdown happened when the naïve Born rule would otherwise predict exceedingly small probabilities.

Reference [2] derived the following expression for the wavefunction of an alpha particle in the laboratory-scale vicinity of an atomic nucleus undergoing slow single-particle s-wave decay:

$$\psi_\alpha(x,t) = \frac{1}{r}\left(\frac{\gamma}{4\pi v}\right)^{1/2} exp\left(\left(\frac{r}{v} - t\right)\left(\frac{\gamma}{2} + i\frac{pv}{\hbar}\right)\right), \tag{1}$$

where $t$ is time, $r$ is distance from the nucleus, $\gamma$ is the decay e-folding rate, $v$ is alpha speed and $p$ is alpha momentum. [This corrects a sign error in [2].] From this I derived the following condition for a visible track to originate at a specific location and time in a cloud chamber, assuming there is an appropriate vapor cluster at the location and time in question:

$$A_\alpha v \tau |\psi_\alpha|^2 > R_c - R, \tag{2}$$

where $\tau$ is the cluster lifetime (presumably due to evaporation), $A_\alpha$ is a parameter that sets the scale for the cross section of an alpha particle to ionize a vapor atom near the threshold of electron energy compensation, $R$ is the radius of the cluster and $R_c$ is critical radius where compensation is exact. If, because of the discreteness of moleular clusters, there is an effective nonzero minimum value for the right-hand side of Equation (2) – call it $\delta_\alpha$ – then there should be a nonzero minimum value of $|\psi_\alpha|^2$ below which a track can't start.

The parameters $A_\alpha$ and $\delta_\alpha$ are very important for the present paper. Our extrapolation to the superconducting qubit will hinge on making assumptions about how these parameters scale to the physics of analog-to-digital (A/D) converters.

The probability density for a track to originate at a specific location and time is related to $|\psi_\alpha|^2$ by

$$\rho \tau A_\alpha v |\psi_\alpha(x,t)|^2, \tag{3}$$

where $\rho$ is the number of vapor clusters formed per unit volume and unit time (multiplied by the numer of ionizing atoms in a cluster), and unit interval of cluster radius $R$. So we can say that there should be a minimum value of naively predicted probability density below which there should be a pronounced deficit of track-start events. Using Equation (1), we can rewrite Equation (3) as

$$\left(\frac{\gamma \rho \tau A_\alpha}{4\pi}\right) e^{-\gamma t} \frac{1}{r^2}. \tag{4}$$

So we can finally say that there should be a maximum value of *r* beyond which there should be a marked deficit of track-start events.

In [3] I observed that, in a cloud chamber experiment with $^{210}$Po alpha decay, there was a deficit of track starts for radii greater than about 2 cm. I identified – and argued against – a number of reasons why this could be experimental artifact. If this actually is a breakdown of the Born rule, then we can work backward to Equation (2), using the parameter estimates in [4], to obtain

$$\delta_\alpha \sim 6 \times 10^{-32} \tau \qquad (5)$$

for $\delta_\alpha$ in meters and $\tau$ in seconds. I do not have a robust theory of $\tau$, but can posit some reasonable bounds. At the lower end, $\tau$ ought to be no smaller than roughly the mean time between cluster encounters with air molecules. The mean free path of an air molecule is ~60nm [7], and the mean speed of an air molecule is ~500m/s; so the mean free time for an air molecule is ~$10^{-10}$s. Since a critical vapor cluster has ~25~$3^3$ molecules [4], the mean time between cluster encounters with air molecules should be mean free time divided by ~$3^2$, or ~$10^{-11}$s. (This is consistent with calculations that yield tens of ps for evaporation of similar-size clusters in a different environment [8] with similar kinetics.) At the higher end, maybe something like 1 sec is not unreasonable since a few seconds is characteristic for small clusters of meteorological interest [9]. Thus we have

$$\sim 10^{-32} \text{Å} < \delta_\alpha < \sim 10^{-21} \text{Å} \qquad (6)$$

These are strikingly small numbers. It seems to me that anything this small compared to the natural scale (Angstrom) of the problem can only be some kind of tunneling effect, but at this time I cannot pinpoint the specific mechanism.

### 3. Extrapolating to superconducting qubit

A superconducting qubit is an artificial atom made from Josephson junctions coupled to a radio frequency (RF) resonating cavity, configured so that the lowest two excited states are close in energy and can be treated together as a two-level system [6]. The state of this two-level system is measured by sending a microwave pure tone at the RF cavity via a transmission line, and recording the reflected signal. If the frequency of the pure tone is chosen appropriately, the signal reflected from one qubit basis state (up or down) has a phase shift that is detectably different from the phase shift due to reflection from the orthogonal basis state. The reflected signal goes through several stages of analog amplification and then passes through an A/D converter, after which it is recorded as a digitized voltage time series. The phase shift is extracted from the time series via conventional I/Q processing, and the measured state is inferred directly from the result. I argued in [5] that canonical quantum measurement behavior per se is localized specifically at the A/D converter, which effectively amounts to an embedded Stern-Gerlach experiment.

In [5], I idealized the A/D converter as a single slab consisting of a cathode layer sandwiched between two semiconductor layers. A real A/D converter [11] consists of many such slabs, each constituting a separate voltage divider, but in the present paper I will continue the single-slab idealization for simplicity. A typical high-performance A/D comes in a square package with side a few mm [12]. For simplicity I assume here that the idealized slab is a square of side 1 cm and therefore area $a=10^{-4} m^2$.

In [5] I accounted for the Born rule in the idealized A/D converter by drawing an analogy between the A/D cathode layer on the one hand, and the mica window of a Geiger counter on the other. To support the analogy, I performed an experiment that strongly suggested that random irregularities on the outside surface of the mica window collimate an incoming alpha-particle wavefunction, leading to ionization tracks in the Geiger-Muller tube that provoke current avalanches at the tube cathode [5,6]. I hypothesized that these window-surface – and therefore A/D cathode layer – irregularities are directly analogous to the exceptional cloud-chamber vapor clusters with extremely large ionization cross sections, and that their exceptional cross sections arise from the same induced-polarization physics.

So there should also be a limit to the Born rule in the qubit case because there should again be a limit to how well induced polarization can compensate for the binding energy of an electron that's ejected from the A/D cathode layer. The exceptional surface irregularity is still a molecular stick figure, rather than a droplet of a continuous, indefinitely adjustable medium. Accordingly, following [5], we can write

$$A_e \left[ P_e \int_{-\infty}^{+\infty} |\psi_e|^2 \, dz \right]_{min} = \delta_e, \qquad (7)$$

for the superconducting qubit, where the use of the subscript $e$ instead of $\alpha$ should be self-explanatory; $P_e$, the naïve Born rule probaiblity, is the absolute-square of the complex amplitude of the two-level qubit state of interest; $\psi_e$ is the normalized copy of the electron wavefunction that propagates into the semiconductor layer of interest; $z$ measures distance in the direction of electron travel (voltage gradient); and the subscript $min$ indicates the smallest value for which the Born rule is valid. Since $|\psi_e|^2$ is normalized, we can roughly approximate the integral in Equation (7) by one over the mean cross-sectional area of the wavefunction. Lacking better insight, we approximate that area by the cross-sectional area of the A/D slab itself, i.e. $a$. Thus we have

$$\frac{P_{e,min} A_e}{a} \sim \delta_e. \qquad (8)$$

If we assume the physics that determines $A_e$ and $\delta_e$ is the same as what determines $A_\alpha$ and $\delta_\alpha$ (except that the projectile $e$ has one-half the charge of the projectile $\alpha$), then we can rewrite Equation (8) as

$$\frac{P_{e,min}A_\alpha}{4a} \sim \delta_\alpha \tag{9}$$

and combine with Equation (5) and the estimated value of $A_\alpha$ in [4] to obtain

$$P_{e,min} \sim 10^{-4} a\tau. \tag{10}$$

Using $a \sim (1 \text{ cm})^2$ and the bounds on $\tau$ in the preceding section, this becomes

$$\sim 10^{-19} < P_{e,min} < \sim 10^{-8}. \tag{11}$$

These numbers seem much too small to have a meaningful impact on practical quantum computing, but a more careful discusion appears warranted.

## 4. Concluding remarks

I have derived a rough estimate of where the Born rule could break down for a two-level superconducting qubit with dispersive readout. The estimate extrapolates from publicly available cloud chamber data, and uses microscopic theories of cloud chamber and A/D converter physics. The theory in the latter instance is considerably more impressionistic than in the former, and in both cases I have to draw on rough arithmetic and sometimes manifestly heuristic assumptions to obtain reportable numbers. The estimate also passes through some very small intermediate numbers (Equation (6)) that seem to call out for an intuitive explanation. I hope this work inspires others to develop much more careful experiments and estimates, even more firmly based on robust microphysics.

It is interesting to reflect philosophically on the strikingly small estimates of the parameter $\delta_\alpha$, which is basically the finest accuracy with which one can talk about the size of an ionizing cluster in a cloud chamber: In avoiding the use of quantum measurement axioms, I am advocating a fully deterministic view of quantum mechanics. This would seem to imply that in principle one could engineer a cloud chamber's population of vapor clusters so that charged particle tracks originated at the same point, measurement trial after measurement trial. M. Pusey has pointed out (private communication) that this could lead to violation of no-signaling theorems [13]. But perhaps that doesn't actually happen because repeatably engineering sub-critical vapor clusters with tolerances as small as in Equation (6) is in principle beyond what quantum-based beings like us can ever do. So maybe the right answer to the most famous Einstein-Bohr dispute is that indeed God doesn't throw dice, but He can exercise a degree of control that we can't even aspire to.

**Acknowledgements**

I am grateful to the organizers of the FQC2025 workshop at the University of Edinburgh, which provided the motivation for this work; and to Matthew Pusey (University of York) for an enlightening conversation.